\documentclass[a4paper,aps,twocolumn]{revtex4}
\RequirePackage[english]{babel}
\RequirePackage[latin1]{inputenc}
\RequirePackage[T1]{fontenc}
\RequirePackage{mathrsfs}
\RequirePackage{amsmath}
\RequirePackage{amssymb}
\RequirePackage{amsbsy}
\RequirePackage{bm}
\begin{document}
\title{\bf{Torsion-Gravity for Dirac fields and their\\ effective phenomenology}}
\author{Luca Fabbri}
\affiliation{INFN \& Dipartimento di Fisica, Universit{\`a} di Bologna,\\
Via Irnerio 46, 40126 Bologna, ITALY}
\date{\today}
\begin{abstract}
We will consider the torsional completion of gravity for a background filled with Dirac matter fields, studying the weak-gravitational non-relativistic approximation, in view of an assessment about their effective phenomenology: we discuss how the torsionally-induced non-linear interactions among fermion fields in this limit are compatible with all experiments, and remarks on the role of torsion to suggest new physics are given.
\end{abstract}
\maketitle
\section*{Introduction and \\motivations}
The torsional completion of gravity is essentially the result of not neglecting the torsion tensor within the most general connection of the spacetime \cite{sa-si,h-o,h}; once torsion is allowed to be non-zero in a geometrical setting in which the metric is non-flat, or equivalently when torsion is allowed to be present beside the curvature tensor, one gets the Cartan enlargement of Riemannian geometry called Cartan-Riemann geometry: such an extension of the underlying structure of the geometrical spacetime might well be justified in terms of generality arguments, but it is in its physical effects that it is most important.

In fact, if gravity is derived as a gauge theory, the gravitational field is interpreted as the strengths of the potentials that arise when making local some continuum spacetime transformation, and so it is all too natural that there be two basic quantities, torsion and curvature, as there are two fundamental spacetime transformations, translations and rotations \cite{Capozziello:2011et, Shapiro:2001rz}: so when gauging the entire Poincar\'{e} group with its full rototranslations, translations give rise to torsion in the same way in which rotations give rise to curvature, according to the approach that was followed by Sciama and Kibble; in the Sciama-Kibble picture torsion turns out to be coupled to the spin in a way that is analogous to the way in which in Einstein gravity curvature is coupled to the energy, so that the Sciama-Kibble scheme is simply the most general expression of the Einsteinian spirit of geometrization of physics, the one in which the spin-torsion coupling is included beside the usual curvature-energy coupling, in what can thus be reasonably called the Sciama-Kibble completion of Einstein gravitation. That this is not only an extension but a true completion comes from the fact that such an enlargement is the most general, first because spacetime rototranslations are all continuum transformations we may gauge and then because spin and energy are all the conserved quantities we may have, according to Wigner classification of elementary particles as irreducible representations of the Poincar\'{e} group; if on the one hand including spin beside energy is the most we can actually do, on the other hand leaving it behind will never permit us to discuss the whole particle content that we know to exist in nature instead. As a consequence of this situation, torsion as well as curvatures are both necessary in order to couple all the conserved quantities of a general matter field precisely because the spin density as well as the energy density are the two conserved quantities that pertain to the most general matter field we may define.

Nevertheless, if from a general point of view we have the prescription for which torsion is coupled to spin and curvature to energy, there is no unique way in which such coupling protocols can indeed be realized unless one fixes additional conditions. For example, what is historically known to be the Sciama-Kibble-Einstein gravitation is based only on the simplest dynamical action one may write, that is the one in which the torsionless Ricci scalar given by $R(g)$ is generalized up to its torsionfull counterpart given by $G(g,Q)$ where $g$ and $Q$ are the metric and torsion tensors; however, because torsion is a tensor then it is not necessary to add it only implicitly though the curvature but it is also possible to add it explicitly as squared torsion contributions \cite{Baekler:2011jt}, and if one relaxes the assumption of parity invariance then also parity-odd terms may be added too \cite{Hojman:1980kv}. And again, these contributions are linear in the curvature and quadratic in torsion, but more terms may be possible if one decides to allow higher-order derivative terms in the Lagrangian of the model: the difference between the aforementioned least-order derivative case \cite{Baekler:2011jt,Hojman:1980kv} and any of the infinite higher-order derivative cases is that while in the former the torsion-spin coupling is algebraic in the latter the torsion-spin coupling equations are differential equations; in the former the torsion-spin coupling is a constrain while in the latter the torsion-spin coupling is a real field equation, so that in the first case torsion is zero when the spin vanishes while in the second case torsion can be present even in vacuum of spin. This property is encoded in the statement that in the least-order derivative models torsion does not propagate out of matter while in all higher-order derivative models it propagates out of matter.

As it is quite clear, all higher-order derivative models may be at risk whenever experimental limits for torsion in vacuum are very strict: this is so because constraining torsion in vacuum means constraining it also inside matter, and therefore totally, with the consequence that torsion might turn out to be zero after all; and because experimental limits for torsion in vacuum are indeed very strict \cite{l,k-r-t, Kostelecky:2008ts}, then either torsion, although non-zero in principle, may be equal to zero because of observational evidence, or there are no higher-order derivative models in the first place. Least-order derivative models are still safe in this case, because torsion is identically zero in vacuum, and then it is compatible with all experimental limits no matter how strict they are outside matter, so that if we wish to have torsion restricted within matter then such constraints can only come from in-matter experiments; very recently \cite{Lehnert:2013jsa}, these experiments have been performed. Our purpose is to discuss what this will imply for the least-order derivative models we consider.

As mentioned above, we will employ the torsional completion of gravity for a geometrical spacetime filled with the most general form of matter accounting for both spin and energy, and so far as we know such a form of matter with both spin and energy is realized in nature by the Dirac field solely; and again, as we have just discussed, we will take the torsion-gravity for Dirac fields as described by Lagrangians at the least-order derivative, in their most general form: the resulting model will be what in this paper we will call Sciama-Kibble-Einstein-Dirac theory, or SKED theory for short. Because the SKED theory is a least-order derivative theory its torsion-spin coupling is algebraic, but because the Dirac field is a least-spin spinorial fermion then it has a completely antisymmetric spin density; this additional feature is important because, since the torsion-spin field equations are algebraic and as the spin is completely antisymmetric, then torsion turns out to be completely antisymmetric itself. What this implies is that, since in least-order derivative models torsion is identically zero in vacuum and therefore it can only be constrained by in-matter experiments, least-order derivative models in which torsion is completely antisymmetric are such that two of the three irreducible decompositions of torsion are identically zero and therefore they cannot be constrained by in-matter experiments however stringent they are, and only the completely antisymmetric irreducible decomposition of torsion can be constrained.

Luckily enough, in the above mentioned in-matter experiments the completely antisymmetric part of torsion is exactly the one on which restrictions are placed, and thus it is essential to perform such an investigation so to assess if torsion will really be constrained and how much.

Eventually, we will speculate about the possible outlooks, especially about hints for new physics.
\section{The SKED Theory}
As specified in the introduction, any higher-order theory of gravity has in general a torsion-spin coupling differential field equation, so that whether or not spin is present, torsion is non-zero in general; in \cite{k-r-t,Kostelecky:2008ts} the authors deal precisely with this type of situation by considering a very general Lagrangian for torsion in interaction with spinorial matter, so that their results are relatively model independent, and capable of including propagating torsion as well: since their results place stringent limits on torsion in vacuum, then they can be interpreted by stating that torsion can be assumed not to exist out of spinorial matter. But in theories in which torsion can propagate, it may be present even in absence of its spin source: then constraining torsion in vacuum signifies constraining torsion entirely, and these results can be interpreted by stating that torsion cannot be a propagating field whatsoever. Or equivalently, since propagating torsion comes from higher-order field equations, they may be interpreted by stating that torsion cannot be described in terms of higher-order Lagrangians in general at all.

Therefore in the present paper we will focus on the least-order Lagrangian, generating torsion-spin coupling algebraic field equations, for which torsion may have whatever value can be assigned in spinorial matter without having to be different from zero also in vacuum, so that torsion may still be present inside matter even if it is always zero outside matter; in \cite{l} the author considers such a theory, so that his results are specific to this model, which describes torsion as a non-linear short-range interaction: effects on the energy levels of atoms can be tested by means of the Hughes-Drever experiment, constraining this specific type of torsion for short-range potentials.

The Lagrangian of \cite{k-r-t,Kostelecky:2008ts} is the starting point also for the results discussed in \cite{Lehnert:2013jsa} although in this last reference the authors discuss in-matter experiments; the results that have been exhibited in \cite{l} about short-range interactions place bounds that in \cite{Lehnert:2013jsa} are improved: therefore we may see the results of \cite{Lehnert:2013jsa} as what condenses and improves all previous results about limits on torsion, placing strong bounds also on the last theory that was still compatible with present experiments, the SKED theory.

Our purpose is to consider the SKED theory, studying how the torsionally-induced spin-contact interactions influence in-matter dynamics, to see whether they really are incompatible with in-matter experiments or not.

So to begin, we will introduce very briefly the formalism we intend to employ, exposed in \cite{Fabbri}, and where here we recall the most important notation: all along this paper we will work in a $(1\!+\!3)$-dimensional space-time with Riemann-Cartan geometry, described in terms of a metric tensor $g_{\mu\nu}$ and a torsion tensor $Q^{\alpha}_{\phantom{\alpha}\mu\nu}$ which will be taken to be completely antisymmetric without any loss of generality for the reasons that were explained in the introduction here above; the metric and torsion tensors will construct the connection in terms of which we define the covariant derivatives $D_{\mu}$ and $\nabla_{\mu}$ in the most general case and in the torsionless case, respectively, and where we have that metric-compatibility holds; then the curvature tensors $G^{\rho}_{\phantom{\rho}\xi\mu\nu}$ and $R^{\rho}_{\phantom{\rho}\xi\mu\nu}$ are defined as usually done in the most general case and in the torsionless case, respectively, and because of their symmetry properties we may also define $G^{\rho}_{\phantom{\rho}\mu\rho\nu}\!=\!G_{\mu\nu}$ with contraction given in terms of $G_{\eta\nu}g^{\eta\nu}\!=\!G$ and $R^{\rho}_{\phantom{\rho}\mu\rho\nu}\!=\!R_{\mu\nu}$ with contraction given by $R_{\eta\nu}g^{\eta\nu}\!=\!R$ called Ricci tensor and scalar and torsionless Ricci tensor and scalar. In Lorentz formalism, the metric is $g_{\alpha\nu}\!=\!e_{\alpha}^{p} e_{\nu}^{i} \eta_{pi}$ in terms of the basis of tetrad fields  $e_{\alpha}^{i}$ and the constant metric $\eta_{ij}$ with Minkowskian structure and where $\omega^{ip}_{\phantom{ip}\alpha}$ is the spin-connection; this formalism is equivalent to the previous one, but it allows the possibility to introduce spinor fields. Here, the spinorial transformation will be taken in $\frac{1}{2}$-spin representation, obtained after introduction of the $\boldsymbol{\gamma}_{a}$ matrices verifying the Clifford algebra $\{\boldsymbol{\gamma}_{a},\boldsymbol{\gamma}_{b}\}\!=\!
2\boldsymbol{\mathbb{I}}\eta_{ab}$ from which one may define the matrices $\frac{1}{4}[\boldsymbol{\gamma}_{a},\boldsymbol{\gamma}_{b}]\!=\!\boldsymbol{\sigma}_{ab}$ such as they verify the condition $\{\boldsymbol{\gamma}_{i},\boldsymbol{\sigma}_{jk}\}\!=\!i\varepsilon_{ijkq} 
\boldsymbol{\pi}\boldsymbol{\gamma}^{q}$ implicitly defining the matrix $\boldsymbol{\pi}$ and where the matrices $\boldsymbol{\sigma}_{ij}$ are the infinitesimal generators of the spinorial transformation, while the spinorial connection $\boldsymbol{\Omega}_{\rho}\!=\! 
\frac{1}{2}\omega^{ij}_{\phantom{ij}\rho}\boldsymbol{\sigma}_{ij}$ defines spinorial covariant derivatives $\boldsymbol{D}_{\rho}$ and $\boldsymbol{\nabla}_{\rho}$ in the general and torsionless case, respectively, thus accomplishing the list of conventions we wanted to recall for the sake of clarity. 

With this kinematic background, we proceed by defining the most general least-order derivative Lagrangian
\begin{eqnarray}
\nonumber
&L\!=\!(\frac{k-1}{4k})Q_{\alpha\nu\sigma}Q^{\alpha\nu\sigma}\!+\!G-\\
&-\frac{i}{2}(\overline{\psi}\boldsymbol{\gamma}^{\mu}\boldsymbol{D}_{\mu}\psi
\!-\!\boldsymbol{D}_{\mu}\overline{\psi}\boldsymbol{\gamma}^{\mu}\psi)\!+\!m\overline{\psi}\psi
\label{actionleast}
\end{eqnarray}
where $k$ is the torsional constant while the gravitational constant has been normalized to unity, and $m$ is the mass of the matter field: in the most general circumstance, torsion enters not only implicitly within the curvature but also explicitly as a quadratic term, both instances having their own coupling constant. In the most general case, as the torsion-squared term is independent from the linear curvature term, the torsional coupling constant is independent from the gravitational Newton constant.

Variation of this Lagrangian with respect to all fields involved yields the corresponding field equations, starting from the completely antisymmetric torsion-spin coupling field equations that are given in the following form
\begin{eqnarray}
&Q^{\rho\mu\nu}\!=\!-k\frac{i}{4}
\overline{\psi}\{\boldsymbol{\gamma}^{\rho}\!,\!\boldsymbol{\sigma}^{\mu\nu}\}\psi
\label{torsion-spin}
\end{eqnarray}
which come together with the non-symmetric curvature-energy coupling field equations given according to
\begin{eqnarray}
\nonumber
&\left(\frac{1-k}{2k}\right)(D_{\mu}Q^{\mu\rho\alpha}
\!-\!\frac{1}{2}Q^{\theta\sigma\rho}Q_{\theta\sigma}^{\phantom{\theta\sigma}\alpha}
\!+\!\frac{1}{4}Q^{\theta\sigma\pi}Q_{\theta\sigma\pi}g^{\rho\alpha})+\\
&+(G^{\rho\alpha}\!-\!\frac{1}{2}Gg^{\rho\alpha})\!=\!\frac{i}{4}(\overline{\psi}\boldsymbol{\gamma}^{\rho}\!\boldsymbol{D}^{\alpha}\psi
\!-\!\boldsymbol{D}^{\alpha}\overline{\psi}\!\boldsymbol{\gamma}^{\rho}\psi)
\label{curvature-energy}
\end{eqnarray}
complemented by the fermionic field equations
\begin{eqnarray}
&i\boldsymbol{\gamma}^{\mu}\!\boldsymbol{D}_{\mu}\psi\!-\!m\psi\!=\!0
\label{fermionic}
\end{eqnarray}
as the most general system of field equations given in terms of the torsional coupling constant and the mass of the matter field as the only unknown parameters.

As anticipated, the assumption of having torsion completely antisymmetric does not require a loss of generality since the spin is completely antisymmetric and the torsion-spin coupling is algebraic; this circumstance also allows us, after all torsionfull quantities have been decomposed in terms of the corresponding torsionless quantities plus torsional contributions, to employ such torsion-spin coupling equations in order to have torsion substituted in terms of the spin of the spinor matter fields, either within the Lagrangian or within the field equations.

When this is done in the Lagrangian we get
\begin{eqnarray}
\nonumber
&L=R\!-\!\frac{i}{2}(\overline{\psi}\boldsymbol{\gamma}^{\mu}\boldsymbol{\nabla}_{\mu}\psi
\!-\!\boldsymbol{\nabla}_{\mu}\overline{\psi}\boldsymbol{\gamma}^{\mu}\psi)-\\
&-\frac{3k}{32}\overline{\psi}\boldsymbol{\gamma}_{\mu}\boldsymbol{\pi}\psi
\overline{\psi}\boldsymbol{\gamma}^{\mu}\boldsymbol{\pi}\psi\!+\!m\overline{\psi}\psi
\label{actionleastdecomposed}
\end{eqnarray}
whose variation will yield the system of gravitational and material field equations already in the decomposed form.

So by varying this action with respect to the metric tensor and the Dirac field or equivalently by substituting torsion in terms of the spin of fermionic fields, we get the symmetric curvature-energy coupling field equations usually known with the name of Einstein field equations
\begin{eqnarray}
\nonumber
&R^{\rho\alpha}\!-\!\frac{1}{2}Rg^{\rho\alpha}
\!=\!\frac{i}{8}(\overline{\psi}\boldsymbol{\gamma}^{\rho}\boldsymbol{\nabla}^{\alpha}\psi
\!-\!\boldsymbol{\nabla}^{\alpha}\overline{\psi}\boldsymbol{\gamma}^{\rho}\psi+\\
\nonumber
&+\overline{\psi}\boldsymbol{\gamma}^{\alpha}\boldsymbol{\nabla}^{\rho}\psi
\!-\!\boldsymbol{\nabla}^{\rho}\overline{\psi}\boldsymbol{\gamma}^{\alpha}\psi)+\\
&+\frac{3k}{64}\overline{\psi}\boldsymbol{\gamma}_{\mu}\boldsymbol{\pi}\psi
\overline{\psi}\boldsymbol{\gamma}^{\mu}\boldsymbol{\pi}\psi g^{\alpha\rho}
\label{gravitational}
\end{eqnarray}
together with the Dirac field equations
\begin{eqnarray}
&i\boldsymbol{\gamma}^{\mu}\boldsymbol{\nabla}_{\mu}\psi
\!+\!\frac{3k}{16}\overline{\psi}\boldsymbol{\gamma}_{\rho}\boldsymbol{\pi}\psi
\boldsymbol{\gamma}^{\rho}\boldsymbol{\pi}\psi\!-\!m\psi\!=\!0
\label{fermionical}
\end{eqnarray}
as the most general system of field equations with torsion replaced by spin-spin contact fermionic interactions of the Nambu--Jona-Lasinio structure and in which the torsional coupling constant has the role of coupling constant giving the strength of these interactions.

By applying to the Dirac equation another Dirac operator and taking advantage of some Fierz rearrangement, it is possible to obtain a Klein-Gordon field equation
\begin{eqnarray}
\nonumber
&\boldsymbol{\nabla}^{2}\psi\!+\!\frac{3k}{8}
\overline{\psi}\boldsymbol{\gamma}^{\mu}\psi i\boldsymbol{\nabla}_{\mu}\psi
\!+\!\frac{3k}{8}\boldsymbol{\nabla}_{\mu}(\overline{\psi}\boldsymbol{\gamma}_{\rho}\psi)
i\boldsymbol{\sigma}^{\mu\rho}\psi-\\
&-\frac{3k}{16}\left(\frac{3k}{16}\!+\!\frac{1}{8}\right) \overline{\psi}\boldsymbol{\gamma}_{\rho}\psi\overline{\psi}\boldsymbol{\gamma}^{\rho}\psi\psi
\!+\!\frac{1}{8}m\overline{\psi}\psi\psi\!+\!m^{2}\psi=0
\end{eqnarray}
in which we notice that even in absence of torsion, encoded by assuming $k$ null, gravitationally-induced non-linear terms are present, rendering non-trivial the dynamics of spinor fields; on the other hand, if we keep torsion while neglecting gravity, then we may take $k$ to be much larger then unity, so that we have the following
\begin{eqnarray}
\nonumber
&\boldsymbol{\nabla}^{2}\psi\!+\!\frac{3k}{8}
\overline{\psi}\boldsymbol{\gamma}^{\mu}\psi i\boldsymbol{\nabla}_{\mu}\psi
\!+\!\frac{3k}{8}\boldsymbol{\nabla}_{\mu}(\overline{\psi}\boldsymbol{\gamma}_{\rho}\psi)
i\boldsymbol{\sigma}^{\mu\rho}\psi-\\
&-\frac{9k^{2}}{256}\overline{\psi}\boldsymbol{\gamma}_{\rho}\psi \overline{\psi}\boldsymbol{\gamma}^{\rho}\psi\psi
\!+\!\frac{1}{8}m\overline{\psi}\psi\psi\!+\!m^{2}\psi=0
\end{eqnarray}
and showing that the non-linearity given by torsion is much more relevant: in this weak-gravity limit we may take stationary configurations of energy $E$ subject to the low-speed regime $E^{2}\!-\!m^{2}\!\approx\!2m(E\!-\!m)$ and hence by writing everything in standard representation we have that the non-relativistic approximation is accomplished by the condition $\overline{\psi} \!\approx\!(\phi^{\dagger},0)$ in terms of which we get
\begin{eqnarray}
\nonumber
&\frac{1}{2m}\!\!\vec{\boldsymbol{\nabla}}\!\cdot\!\vec{\boldsymbol{\nabla}}\phi
\!+\!\frac{9k^{2}}{512m}|\phi^{\dagger}\phi|^{2}\phi-\\
&-\frac{1}{16}|\phi^{\dagger}\phi|\phi\!+\!(E\!-\!m)\phi\!=\!0
\end{eqnarray}
as Pauli-Schr\"{o}dinger field equations for a non-relativistic semi-spinor matter field. But on the other hand, we also have to notice that the absence of the Pauli matrices means that the Pauli-Schr\"{o}dinger field equations decouple in one Schr\"{o}dinger field equation for each of the two components of the semi-spinor field that can then be taken as independent, and so we actually have
\begin{eqnarray}
\nonumber
&\frac{1}{2m}\!\!\vec{\boldsymbol{\nabla}}\!\cdot\!\vec{\boldsymbol{\nabla}}u
\!+\!\frac{9k^{2}}{512m}|u^{*}u|^{2}u-\\
&-\frac{1}{16}|u^{*}u|u\!+\!(E\!-\!m)u\!=\!0
\label{equation}
\end{eqnarray}
as a Schr\"{o}dinger field equation for a non-relativistic complex scalar field: the absence of Pauli terms in the non-linearities for the single Dirac field encodes the fact that there is a complete isotropy in the self-interaction of the single matter field. Such self-interactions cannot be detected in the type of experiments we are considering.

If in the Lagrangian beside the initial matter field we were to include a second matter field then we would get
\begin{eqnarray}
\nonumber
&L\!=\!(\frac{k-1}{4k})Q_{\alpha\nu\sigma}Q^{\alpha\nu\sigma}\!+\!G-\\
\nonumber
&-\frac{i}{2}(\overline{\psi}\boldsymbol{\gamma}^{\mu}\boldsymbol{D}_{\mu}\psi
\!-\!\boldsymbol{D}_{\mu}\overline{\psi}\boldsymbol{\gamma}^{\mu}\psi)-\\
&-\frac{i}{2}(\overline{\chi}\boldsymbol{\gamma}^{\mu}\boldsymbol{D}_{\mu}\chi
\!-\!\boldsymbol{D}_{\mu}\overline{\chi}\boldsymbol{\gamma}^{\mu}\chi)
\!+\!m\overline{\psi}\psi\!+\!M\overline{\chi}\chi
\end{eqnarray}
so that torsion would now be given as the total spin, accounting for both the initial fermion and the supplementary fermion; the effective Lagrangian is thus
\begin{eqnarray}
\nonumber
&L=R\!-\!\frac{i}{2}(\overline{\psi}\boldsymbol{\gamma}^{\mu}\boldsymbol{\nabla}_{\mu}\psi
\!-\!\boldsymbol{\nabla}_{\mu}\overline{\psi}\boldsymbol{\gamma}^{\mu}\psi)-\\
\nonumber
&-\frac{i}{2}(\overline{\chi}\boldsymbol{\gamma}^{\mu}\boldsymbol{\nabla}_{\mu}\chi
\!-\!\boldsymbol{\nabla}_{\mu}\overline{\chi}\boldsymbol{\gamma}^{\mu}\chi)-\\
\nonumber
&-\frac{3k}{16}\overline{\psi}\boldsymbol{\gamma}_{\mu}\boldsymbol{\pi}\psi
\overline{\chi}\boldsymbol{\gamma}^{\mu}\boldsymbol{\pi}\chi
\!-\!\frac{3k}{32}\overline{\psi}\boldsymbol{\gamma}_{\mu}\boldsymbol{\pi}\psi
\overline{\psi}\boldsymbol{\gamma}^{\mu}\boldsymbol{\pi}\psi-\\
&-\frac{3k}{32}\overline{\chi}\boldsymbol{\gamma}_{\mu}\boldsymbol{\pi}\chi
\overline{\chi}\boldsymbol{\gamma}^{\mu}\boldsymbol{\pi}\chi
\!+\!m\overline{\psi}\psi\!+\!M\overline{\chi}\chi
\end{eqnarray}
perfectly symmetric in the two fermions: however, if the initial fermion is kept as dynamical while the second is taken as fixed, the effective Lagrangian is simply
\begin{eqnarray}
\nonumber
&L=R\!-\!\frac{i}{2}(\overline{\psi}\boldsymbol{\gamma}^{\mu}\boldsymbol{\nabla}_{\mu}\psi
\!-\!\boldsymbol{\nabla}_{\mu}\overline{\psi}\boldsymbol{\gamma}^{\mu}\psi)-\\
&-\frac{3k}{16}\overline{\psi}\boldsymbol{\gamma}_{\mu}\boldsymbol{\pi}\psi
\overline{\chi}\boldsymbol{\gamma}^{\mu}\boldsymbol{\pi}\chi
\!-\!\frac{3k}{32}\overline{\psi}\boldsymbol{\gamma}_{\mu}\boldsymbol{\pi}\psi
\overline{\psi}\boldsymbol{\gamma}^{\mu}\boldsymbol{\pi}\psi
\!+\!m\overline{\psi}\psi
\end{eqnarray}
in which as it is straightforward to see an asymmetry has appeared among the two fields. With this Lagrangian we describe the dynamics of the initial fermion in self-interaction and in interaction with the additional fermion taken to represent a non-dynamical background, and because of the fact that the self-interactions cannot be detected in the type of experiments we are considering then
\begin{eqnarray}
\nonumber
&L=R\!-\!\frac{i}{2}(\overline{\psi}\boldsymbol{\gamma}^{\mu}\boldsymbol{\nabla}_{\mu}\psi
\!-\!\boldsymbol{\nabla}_{\mu}\overline{\psi}\boldsymbol{\gamma}^{\mu}\psi)-\\
&-\frac{3k}{16}\overline{\psi}\boldsymbol{\gamma}_{\mu}\boldsymbol{\pi}\psi
\overline{\chi}\boldsymbol{\gamma}^{\mu}\boldsymbol{\pi}\chi
\!+\!m\overline{\psi}\psi
\end{eqnarray}
will provide the same amount of information about the observables of the system. And this is the Lagrangian we will investigate in the following of the paper.

This Lagrangian is a generalization of the Lagrangian that was obtained in \cite{Fabbri}; in the following it will be compared to the results obtained in \cite{Lehnert:2013jsa}: the idea is to assess the way in which a model described by such a Lagrangian can be such that the completely antisymmetric part of torsion could escape from the constraints imposed in non-relativistic regimes by in-matter experiments.

To do that, we consider that in \cite{Lehnert:2013jsa}, the authors start from a model-independent Lagrangian that nevertheless has to be applied to the case of polarized slow neutrons through a condensed-state of liquid $^{4}\mathrm{He}$ taken as a background field distribution: first of all, as it has been discussed in the above reference, the geometrical properties of such a system are such that the torsional effects have to be isotropic, and the authors go ahead in explaining what restrictions on the parameters are allowed, so to simplify the Lagrangian function; then, because all fermions involved are Dirac fields, their spin is completely antisymmetric and thus they can only generate a torsion that is completely antisymmetric, its dual indicated in terms of the fixed axial vector $A_{\mu}$ to follow the notation of the above paper, and this also amounts to additional simplifications in the Lagrangian; finally, as we intend to compare this Lagrangian to the one we have in the present model, the highest-order derivative are to be present in the kinetic term alone, for further simplification in their Lagrangian once applied to the SKED theory: when all these requirements are implemented, all parameters can be taken to vanish except a single one, so that
\begin{eqnarray}
\nonumber
&L=-\frac{i}{2}(\overline{\psi}\boldsymbol{\gamma}^{\mu}\boldsymbol{\nabla}_{\mu}\psi
\!-\!\boldsymbol{\nabla}_{\mu}\overline{\psi}\boldsymbol{\gamma}^{\mu}\psi)+\\
&+\xi_{4}^{(4)}A_{\mu}\overline{\psi}\boldsymbol{\gamma}^{\mu}\boldsymbol{\pi}\psi\!+\!m\overline{\psi}\psi
\end{eqnarray}
in terms of the single $\xi_{4}^{(4)}$ parameter is the Lagrangian we will have to employ to fit our model; this Lagrangian gives rise to the non-relativistic Hamiltonian given by
\begin{eqnarray}
&H\!\approx\!\frac{P^{2}}{2m}
\!+\!\vec{\frac{P}{m}}\!\cdot\!\vec{\frac{\boldsymbol{\sigma}}{2}}(-2\xi_{4}^{(4)}A^{0})
\label{function}
\end{eqnarray}
placing bounds on the $-2\xi_{4}^{(4)}A^{0}$ term. It is worth noticing that not the individual factors but only the entire term will be constrained by experimental measurements.

We recall that such Hamiltonian is the one describing the dynamics of the slow neutrons in a background in which the torsion is generated inside the liquid $^{4}\mathrm{He}$.

We are now able to compare the results, knowing that what in our model was the initial fermion is identified with the neutron while the additional fermion is identified with the liquid $^{4}\mathrm{He}$, and every neutron has self-interactions and interactions with liquid $^{4}\mathrm{He}$ in the most general circumstances: however, the complete isotropy of the self-interaction for a single matter field means that the contribution given by $\overline{\psi}\boldsymbol{\gamma}_{\mu}\boldsymbol{\pi}\psi \overline{\psi}\boldsymbol{\gamma}^{\mu}\boldsymbol{\pi}\psi$ is not going to give rise to any correction to the Hamiltonian in the form that is given in expression (\ref{function}), so that we may neglect the self-interaction for each neutron field, and thus
\begin{eqnarray}
&-\frac{3k}{16}\overline{\chi}\boldsymbol{\gamma}^{\mu}\boldsymbol{\pi}\chi
\!=\!\xi_{4}^{(4)}A^{\mu}
\end{eqnarray}
as it is easy to check; in the standard representation writing the fermion according to $\overline{\chi}\!\approx\!(a^{\dagger},-b^{\dagger})$ we finally have
\begin{eqnarray}
&\frac{3}{8}k(a^{\dagger}b\!+\!b^{\dagger}a)\!=\!-2\xi_{4}^{(4)}A^{0}
\end{eqnarray}
and the bounds on $-2\xi_{4}^{(4)}A^{0}$ are on $k(a^{\dagger}b\!+\!b^{\dagger}a)$ instead, but in standard representation $b$ is the small-valued semi-spinorial component, the one that vanishes in the non-relativistic limit. Such limit is certainly applicable in this case since the liquid $^{4}\mathrm{He}$ is static, with the consequence that the mixed term $k(a^{\dagger}b 
\!+\!b^{\dagger}a)$ vanishes, because of the vanishing of the field, and therefore showing that this term is compatible, regardless the actual value of the constant $k$, with any constraint placed by the experiment.

As a consequence of this fact, we have that the SKED theory we have introduced above is the only gravitational theory which, even in its most general instance given when the torsional coupling constant is completely undetermined, is compatible with all in-matter experiments.

Finally we will discuss some of the consequences.
\section{Effective Interactions}
Up to now we discussed that the theory in \cite{Fabbri} is among all the least-order derivative dynamical theories the one that is the most general coupling the torsional completion of gravity to spinorial matter fields: it gives Dirac matter field equations with non-linear potentials in which the torsional coupling constant is not determined by any empirical results; in particular its non-relativistic limit results into Schr\"{o}dinger field equations that contain no Pauli contribution. This situation holds especially for polarized slow neutrons; experiments such as those involving polarized slow neutrons in interaction with a condensed-state of liquid helium, which is static, will have no Pauli term as a correction to the Hamiltonian of the system, and therefore the system is not constrained by any measurement. The SKED theory remains the only gravitational theory compatible with observations.

To probe torsionally-induced non-linear interactions one is compelled to study in-matter experiments performed in the relativistic regimes, involving the high-energy scattering of many particles: these high-energy scattering can probe models up to a few TeV solely, for the moment being. Thus the spectrum ranging from the present to the Planck scale is still largely unbound.

Nevertheless, this opens an interesting question about torsion, but before dealing with that, we would like to spend some words in order to clarify a misconception that is unfortunately quite widespread: the torsional completion of gravity is achieved by not neglecting torsion beside curvature, as the two fundamental objects describing the character of the spacetime; however, because torsion is a tensor on its own, the action should not only have a curvature tensor implicitly containing torsion but torsion should also be explicitly present in terms of squared contributions, thus accounting for an independent constant that is different from the gravitational constant in general circumstances. Because Einstein gravity can be obtained variationally from the Lagrangian that is given by the Ricci torsionless curvature scalar $R$ people initially obtained the torsional completion of Einstein gravity from the variation of the Lagrangian that is given by the Ricci torsionfull curvature scalar $G$, but then an action containing only $G$ has only one term, and therefore it cannot have more than one constant, which must be nothing else but the Newton constant in order to recover Newtonian dynamics in the weak-gravitational low-speed static configurations: overlooking the fact that more general Lagrangians were possible has laid the basis for the misconception that the torsional constant had to be the Newton constant, and such a misconception was eventually cemented along the decades. Hence we would like to take the opportunity here to clearly stress the fact that the Lagrangian given by $G$ is certainly the most straightforward but nevertheless not the most general Lagrangian, which is given by the Ricci torsionfull curvature scalar accompanied by quadratic torsion terms, therefore given in terms of two different constants, the gravitational one being the Newton constant but the torsional one being completely undetermined. As a consequence of the fact that the torsional constant might be much larger than the Newton constant, we have that the torsionally-induced non-linear terms within the Dirac matter field equations might be relevant much before the Planck scale.

As a matter of fact, it may happen that the torsional constant is not much larger than the Newton constant, and the torsionally-induced non-linear terms of the Dirac matter field equations are relevant only at the Planck scale after all, but this is not a necessity; now back to the problem of the boundaries on torsion, we have just recalled that, on the other hand of the allowed spectrum, the torsional constant cannot be larger than the Fermi constant, or else the torsionally-induced non-linear terms of the Dirac matter field equations would have been relevant before the Higgs scale, but we have never detected them at those distances: this places the torsional constant between the Newton and Fermi constant, so not much of a constraint. The torsional constant must be smaller than the Fermi constant, but because it does not need to be as small as the Newton constant, it might happen that the torsion constant is just a little smaller than the Fermi constant: if the torsional constant were just a little smaller than the Fermi constant, then what would the consequences be? Would this be of any help in addressing open problems in physics or in constituting evidence suggesting the appearance of new physics, right beyond what can be probed at the moment?

For instance, in field theory, computing some quantities may lead to divergences unless a cut-off is introduced by hand, and even so it may well happen that a reasonable cut-off may still give exceedingly large results compared to observations; the whole idea of placing a cut-off beyond which computations cannot be done is interpretable by thinking that there is a limit beyond which new effects change the physics in such a way that the same computations done in terms of this new physics would give finite results: a theory with a torsional coupling constant that happens to be just a little smaller than the Fermi constant, so that the torsionally-induced non-linear interactions happen to become relevant a little beyond these scales, does precisely this. If we interpret the torsional coupling constant as the effective limit encoded by the cut-off of the theory and the torsionally-induced non-linear terms as new physics, all computations in the standard context would happen to work properly up to the scale at which there is the cut-off because the torsional effects are negligible, and beyond such scales calculations would no longer be reliable because torsional effects would change the effective phenomenology; if the torsional coupling constant were to be a little smaller than the Fermi constant the cut-off would be just beyond the present scales, and problems related to divergences would not necessarily appear beyond this boundary.

Thus, if the torsional coupling constant happened to be tuned a little beyond the Fermi constant, it would mean that all the torsionally-induced non-linear interactions in high-energy scattering would become manifest soon after the scales we are probing in today's accelerators, with the interesting consequence that such non-linearities might soon let new physics arise; this might be of some help in addressing problems that can be solved when new physic is necessary. Again, it may well be that after all the torsional coupling constant will be measured to be much smaller, no torsionally-induced non-linear term will be relevant and no new physics will be possible along this avenue, but for the moment this is a viable possibility.

Even if we cannot yet be sure that torsional effects will be relevant a little beyond the Fermi scale, nevertheless a situation in which this could happen is better than the situation in which torsional effects were not thought to be possible anywhere before the Planck scale.

More information about such effects may only come from high-energy physics experiments.
\section*{Conclusion}
In this paper, we considered the torsional completion of gravitation in a spacetime filled with Dirac fields, specifying that we had taken least-order derivative dynamics, which we called SKED theory, and we discussed SKED models in two situations: one in which there was a single matter field, for which we have shown that the non-linear potentials were isotropic; and another in which a matter field was sent to probe a non-dynamical static matter field distribution, for which we have shown that the non-linear potentials were vanishing. We have discussed that when such models are used to describe neutrons in interactions with static liquid $^{4}\mathrm{He}$ as in recent in-matter experiments, the non-linear potentials account for either a self-interaction that cannot be detected by such experiments or by mutual interactions that nevertheless are equal to zero, and so all these torsionally-induced non-linear potentials are compatible with all limits that are set by the type of in-matter experiments discussed in the recent literature; furthermore, we have remarked that these results are true regardless the value of the torsional coupling constant. Therefore torsion in gravity for least-spin spinor fields in the least-order derivative action in its most general case is at the same time the simplest and yet the most general theory that is still compatible with all experimental constraints we know at the moment.

In the second part of the paper, we have discussed that the only way we may have to detect the torsional effects is by studying anisotropies in relativistic scattering, commenting that this type of experiments may take place at the LHC, although we have specified that beyond the Fermi scale there is no constraint that has been placed yet; then we went on to discussing that in a situation in which the torsional coupling constant may be anywhere between the Fermi and the Planck scales, such a constant might happen to be just a little smaller than the Fermi constant, and we have stressed that interesting consequences might follow. As in the final part of the paper we have thoroughly specified, all this does not mean that having torsionally-induced non-linear interactions relevant right beyond the Fermi scale is what will actually happen, but at least the possibility of this occurrence may be view as an opportunity to consider new physics right beyond what we observe today.

Then only accelerators may tell.

\end{document}